\documentclass[final,prd,twocolumn,superscriptaddress,groupedaddress,nofootinbib,floatfix]{revtex4}
\usepackage[sort&compress]{natbib}
\usepackage{graphicx}
\usepackage{paralist,array}
\usepackage{units}                    
\graphicspath{{fig/}{./Plots/}} 
\usepackage[dvipsnames,usenames]{color,xcolor} 
\usepackage[colorlinks=true,citecolor=blue]{hyperref}
\newcommand{\GeV}{\:\unit{GeV}}
\newcommand{\gev}{\GeV}

\newcommand{\myparagraph}[1]{\bigskip \noindent\textbf{#1}}
\begin{document}
\title{Features in the Standard Model diphoton background}
\author{Kyrylo~Bondarenko}
\affiliation{Instituut-Lorentz for Theoretical Physics, Universiteit Leiden, Niels Bohrweg 2, Leiden, The Netherlands}

\author{Alexey~Boyarsky}
\affiliation{Instituut-Lorentz for Theoretical Physics, Universiteit Leiden, Niels Bohrweg 2, Leiden, The Netherlands}

\author{Oleg~Ruchayskiy}
\affiliation{Discovery Center, Niels Bohr Institute, Blegdamsvej 17, DK-2100 Copenhagen, Denmark}

\author{Mikhail Shaposhnikov}
\affiliation{Ecole Polytechnique F\'ed\'erale de Lausanne, FSB/IPHYS/LPPC, BSP
  720, CH-1015, Lausanne, Switzerland}
\date{\today}
\begin{abstract}
We argue that the electromagnetic decays of energetic unflavoured neutral mesons, notably $\eta$, mis-identified as single photons due to granularity of the electromagnetic calorimeter  might create  bump-like features in the diphoton invariant mass spectrum at different energies, including 750 GeV.  We discuss what kind of additional analysis can exclude or confirm  this hypothesis.
\end{abstract} 
\maketitle

\section{Introduction}

Recent reports of excess in the diphoton invariant mass spectrum at energies
about 750~GeV~\cite{ATLAS-diphoton,CMS:2015dxe,ATLAS-diphoton2016} have
generated a lot of interest in the
community~\cite{Backovic:2016xno,resonaances}. Most of the works concentrated
on interpretations of the excess as coming from some new particle, while few
explored also the statistical significance of the
signal~\cite{Davis:2016hlw,Buckley:2016mbr,Kavanagh:2016pso}.

The  ``data-driven'' background fits lead to smooth functions at energies of
interest~\cite{ATLAS-diphoton,CMS:2015dxe} (see also~\cite{Aad:2015zhl} in
context of Higgs $\to \gamma\gamma$ searches).  While the functional form of
the backgrounds used for the analysis was challenged in~\cite{Davis:2016hlw},
the diphoton \emph{Standard Model background} is expected to be
\emph{monotonic} across the energies of interest (the recent
work~\cite{Jain:2016kai} discussed top-quark threshold effects at lower energies
$m_{\gamma\gamma} \sim 2 m_{top}$). Studies of the direct (prompt) diphoton
production confirm this expectation
\cite{Binoth:1999qq,Bern:2001df,Bern:2002jx,Li:2011ye,Catani:2011qz,Campbell:2016yrh}.

\bigskip

\begin{figure}
  \centering
  \includegraphics[width=0.5\textwidth]{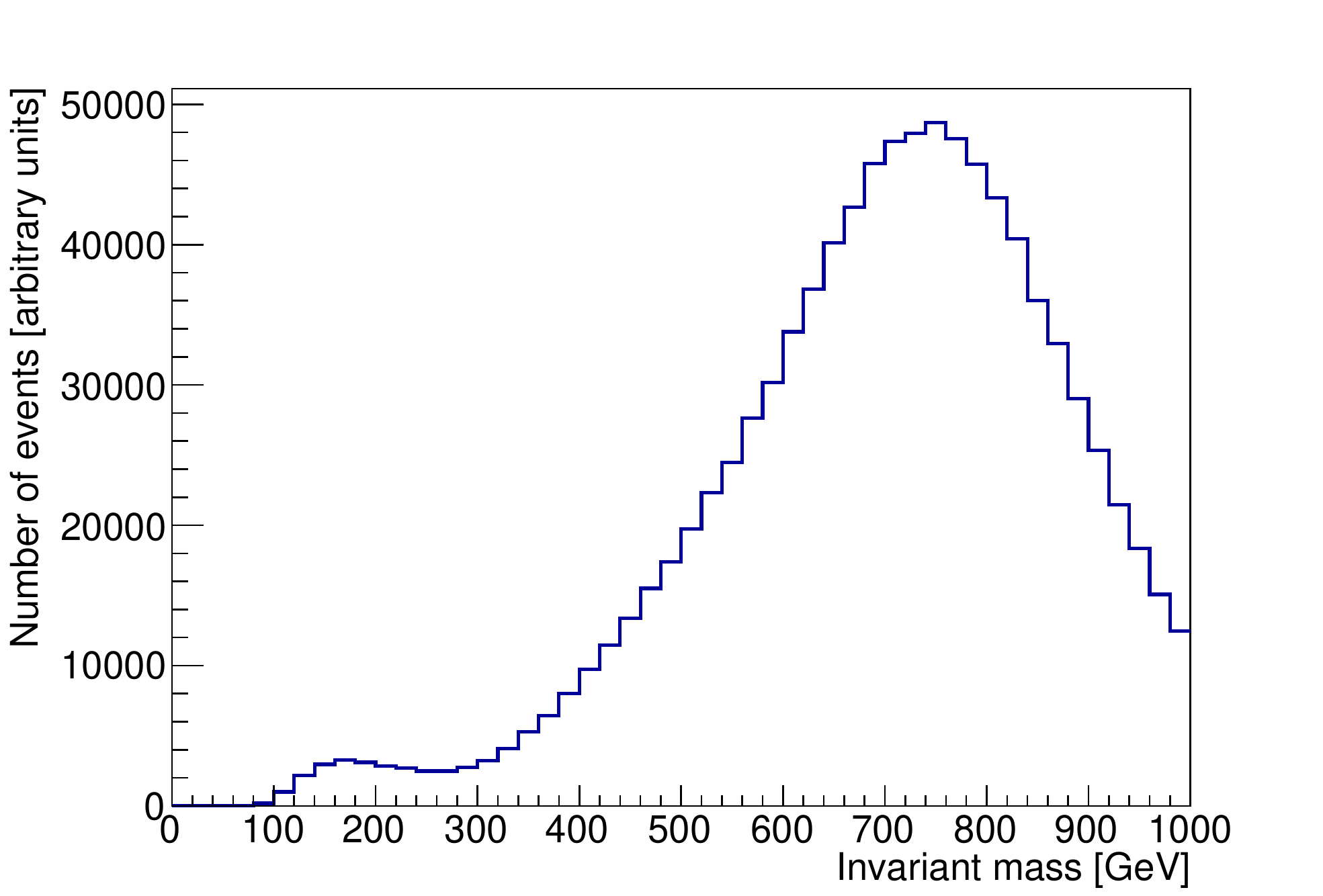}
  \caption{Invariant mass of two $\eta$-mesons mis-identified as photons
    with the ATLAS detector for the $\eta\to 3\pi^0 \to 6\gamma$ decay
    process. See text for details.}
  \label{fig:dieta_mass_Energy100_ATLASsq_3pions_05}
\end{figure}

\emph{This paper scrutinizes the assumption of the
  smoothness of the background.} Our main idea is the following: while the
physical background is indeed smooth, due to a finite
granularity of the electromagnetic calorimeter, sufficiently boosted neutral
mesons (such as $\pi^0$ or $\eta$), decaying electromagnetically, cannot be
distinguished from a single photon, travelling in the direction of the meson
and carrying all its energy\footnote{Such decays are well known as one of the
  main non-prompt backgrounds for photon detection at ATLAS~\cite{Aad:2010sp}
  and CMS~\cite{Khachatryan:2015iwa}. A lot of work has been done to analyze
  this background, in particular in the domain of energies of the Standard
  Model Higgs.}. The probability of such a misinterpretation sharply increases
with energy of the incoming neutral meson, while the overall number of mesons
drops fast with energy.  The convolution of growing and decaying functions
leads to a bump-like feature in the energy distribution of photons and,
correspondingly, propagates to the diphoton spectrum.  The position of this
feature depends on three main factors: energy of incoming particle, size of
the calorimeter's granularity, and the type of the incident neutral meson. 
Going straight to our main result, we find that the bumps which may result
from $\eta\to 3\pi^0\to6\gamma$ decays in the ATLAS
detector can indeed appear around 750 GeV (see
Fig.~\ref{fig:dieta_mass_Energy100_ATLASsq_3pions_05}).

The
idea that some hypothetical neutral particles, decaying to two photons can be
mis-interpreted as a single photon had been previously invoked in the context of $H\to
\gamma\gamma$ process 
\cite{Dobrescu:2000jt,Chang:2006bw,Toro:2012sv,Chang:2008cw,Draper:2012xt,Ellis:2012zp,Ellis:2012sd,Curtin:2013fra}
or $750\gev$ excess
\cite{Knapen:2015dap,Bi:2015lcf,Chang:2015sdy,Agrawal:2015dbf,Chala:2015cev,Aparicio:2016iwr,Chen:2016sck}. However,
to the best of our knowledge the question of  $\eta\to 3\pi^0\to6\gamma$ decays contributing to the diphoton spectrum has not been analysed either theoretically or by using Monte
Carlo simulations combined with the simulated
detector responses.

\section{Photons from neutral meson decays} 

\begin{figure}
  \centering
  \includegraphics[width=0.5\textwidth]{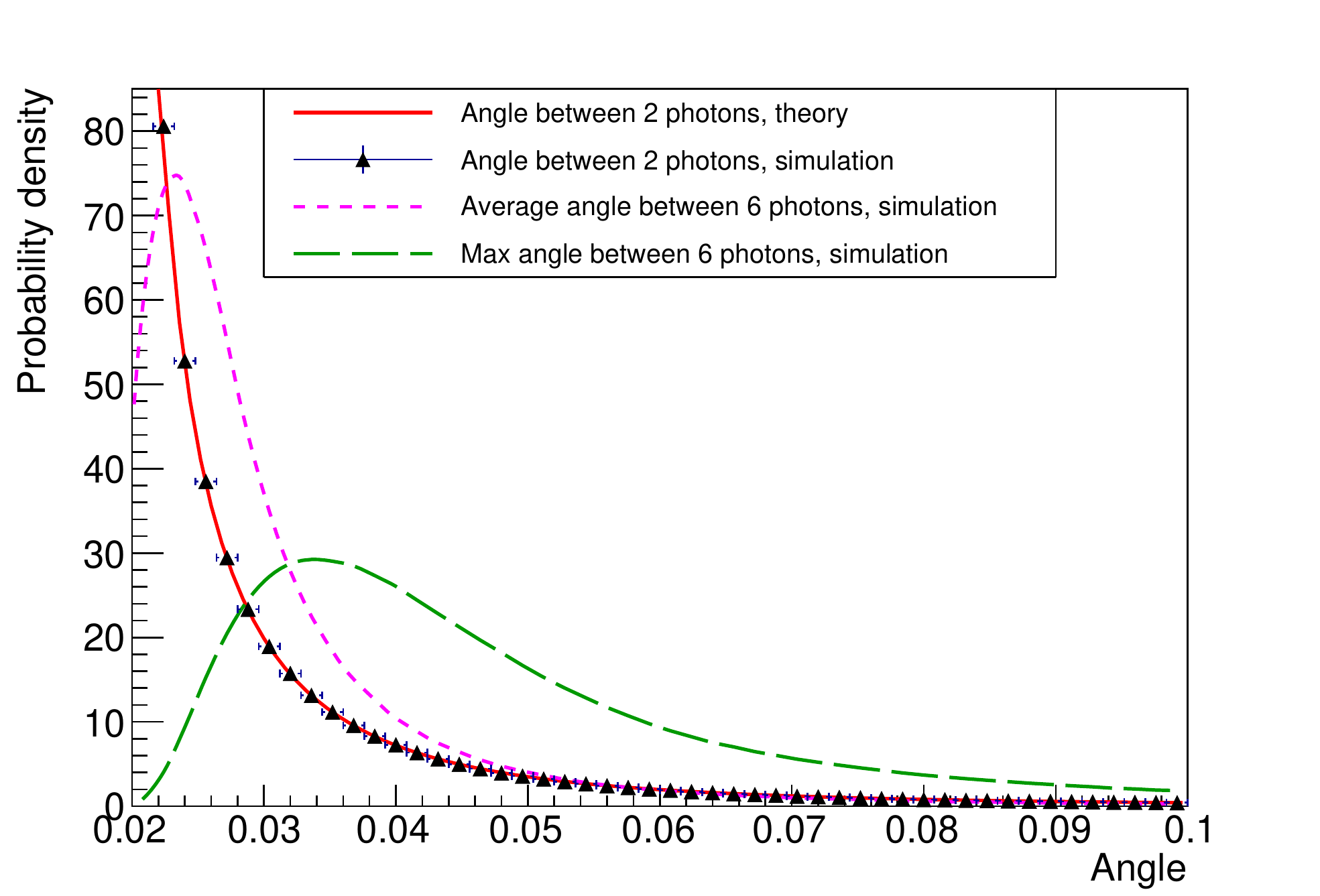}
  \caption{Angular separation of photons from meson decay. Decays
    $\eta \to 2\gamma$: red line is Eq.~(\protect\ref{eq:4}) vs.\ blue (MC)
    point. Green line -- \emph{maximal} angle among 6 photons
    for $\eta \to 6\gamma$. Average angle between 6 photons and the direction
    of original $\eta$-meson is shown in magenta, short-dashed line.}
  \label{fig:angulardistribution}
\end{figure}

Consider neutral unflavoured
mesons, $\pi^0$ and $\eta$, that decay to two photons (branching
ratios: $Br_{\pi^0\to2\gamma} \simeq 100\%$,
$Br_{\eta \to 2\gamma} \approx 39.4\%$~\cite{Agashe:2014kda}).  The
distribution of photons in the meson's rest frame is isotropic, therefore in
the laboratory frame where the meson has Lorentz factor $\gamma$, the
distribution in $\alpha$ -- angle between two photons -- has the form
\begin{equation}
  \label{eq:4}
  \frac{dN_\gamma}{d\alpha} = \frac{1}{2\sqrt{\gamma^2 -1}}\frac{\cos(\alpha/2)}{\sin^2(\alpha/2)}\frac1{\sqrt{\gamma^2 \sin^2(\alpha/2) - 1}}\;,
\end{equation}
where $dN_\gamma$ is the number of photon pairs with the separation in the laboratory frame between $\alpha$ and $\alpha + d\alpha$ (see Fig.~\ref{fig:angulardistribution}). 
The minimal angle between two photons is therefore
\begin{equation}
  \label{eq:1}
  \alpha_{min} = 2\arcsin(\gamma^{-1}) \approx \frac2\gamma \quad\text{for}\quad \gamma \gg 1\;.
\end{equation}
The distribution~(\ref{eq:4}) is sharply peaked and $95\%$ of all photons have
angles $\alpha_{min} < \alpha < 3\alpha_{min}$.

In case of $\eta$-meson, there is another neutral decay mode:
$\eta \to 3\pi^0$ ($Br\simeq 32.7\%$,~\cite{Agashe:2014kda}) with subsequent
decay of each $\pi^0\to2\gamma$ (\emph{for simplicity in what follows we call
  this mode $\eta \to 6\gamma$}). To simulate the distribution of resulting
photons, we wrote a \textsc{root}~\cite{Brun:1997pa} program, that use \textsc{TGenPhaseSpace}
utility to simulate n-body decays. In our analysis we assumed that the $3\pi^0$ angular distribution is isotropic in the $\eta$ rest frame and did-not take into account the energy dependence of the corresponding matrix element \cite{Gasser:1984pr}.

\begin{figure}
  \centering
  \includegraphics[width=0.5\textwidth]{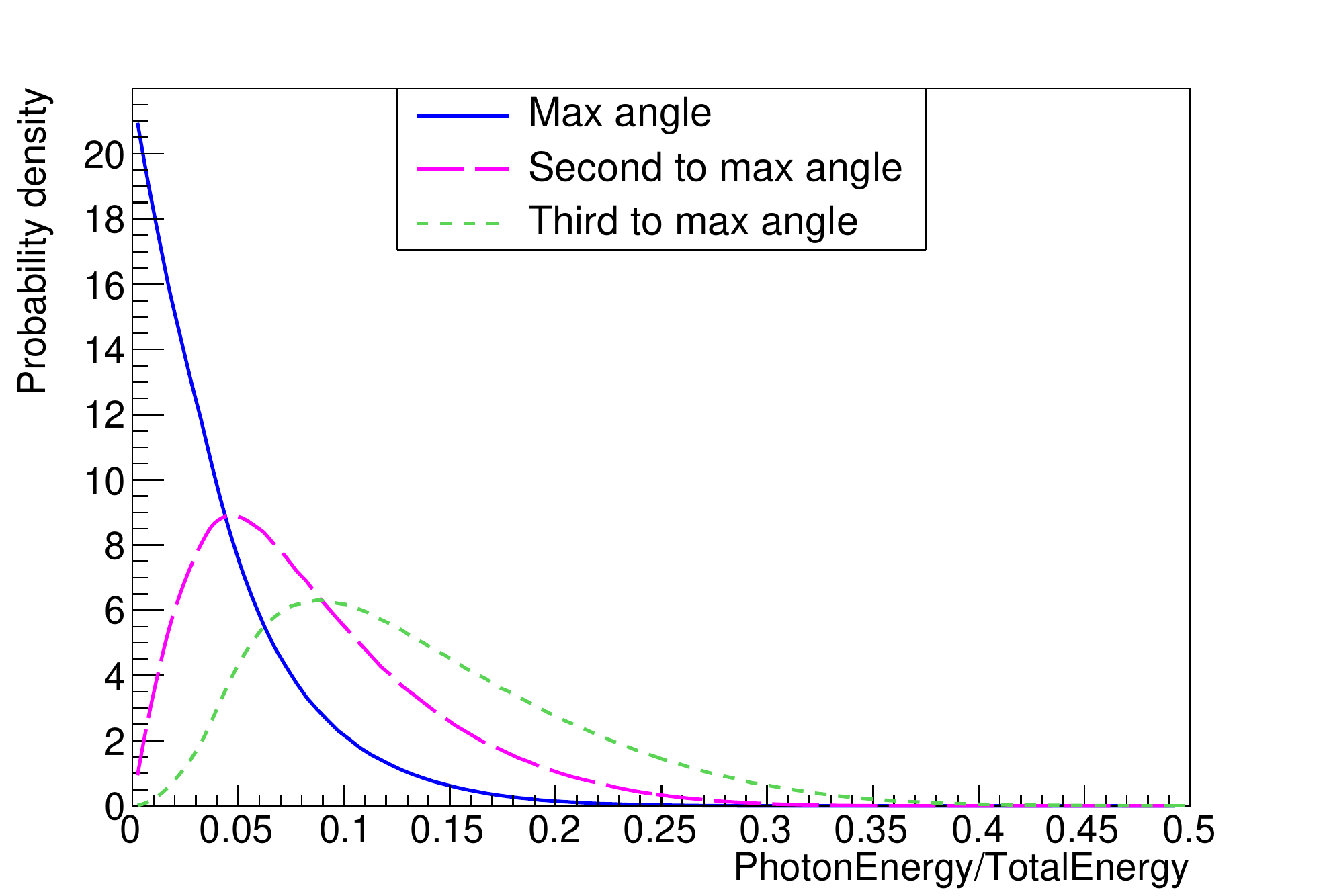}
  \caption{Fraction of $\eta$-meson energy, carried by a photon that is
    maximally away from the original meson's direction, then second and third
    largest angles.}
  \label{fig:EnergyFractionAngle}
\end{figure}

Given that $m_\eta - 3 m_\pi = \unit[143]{GeV}$, $\pi^0$ mesons, arising in
the decay $\eta \to 3\pi^0$, are mildly- or non-relativistic in the $\eta$-meson's rest
frame. Then one should have $\gamma_{\pi^0} \approx \gamma_{\eta}$ and 
could expect that 6 photons arrive collimated similarly to the 2 photon case.
Nevertheless, as the average momentum of $\pi^0$ in the rest frame of the
$\eta$ mesons is $|\vec p_{\pi^0}| \sim \unit[120]{MeV}$, the total width of
the photon distribution (maximal angle between a photon and the direction of
$\eta$-meson) is wider, than in the two photon case (green vs.\ red curve in
Fig.~\ref{fig:angulardistribution}), but not surprisingly, most of the energy
is contained in the photons, that are ``closer than on average'' to the
direction of the original meson (in the rest frame of $\eta$-meson these
photons are emitted closest to the direction of the boost), see dashed magenta
line in Fig.~\ref{fig:angulardistribution}, see also Fig.~\ref{fig:EnergyFractionAngle}.

\begin{table}[!t]
  \centering
  \begin{tabular}{|l|>{$}l<{$}|>{$}l<{$}|l|}
    \hline
    Layer                       & \eta-\text{direction}  & \phi-\text{direction}  & Comment \\
    \hline
    \multicolumn{4}{|c|}{ATLAS} \\
    \hline
    1st layer                   & \Delta \eta_A = 0.003 -
    0.006                       & \Delta\phi_A = 0.1     & $\eta$-dependent       \\
    2nd layer                   & \Delta \eta_A = 0.025  & \Delta\phi_A = 0.025   &         \\ 
    \hline
    \multicolumn{4}{|c|}{CMS}   \\
    \hline
                                & \Delta \eta_C = 0.0174 & \Delta\phi_C  = 0.0174 &         \\
    \hline                                         
  \end{tabular}
  \caption{Granularities of ATLAS and CMS calorimeters.} 
  \label{tab:granulariites}
\end{table}

\section{Meson-photon misidentification} 

Next we take into account that
both ATLAS and CMS electromagnetic calorimeters (ECAL) have finite spatial
resolution. The ATLAS calorimeter is lead-liquid argon sampling calorimeter
with an accordion geometry (described in details e.g.\ in
\cite{Aad:2009wy}). Its characteristics, relevant for our analysis are listed
in Table~\ref{tab:granulariites}. The CMS ECAL is made of $\text{PbWO}_4$
crystals, with square cross-section (Table~\ref{tab:granulariites}).

By considering the 2 photon decay and requiring $\alpha_{min} \le 0.003$
(minimal granularity in $\eta$-direction for ATLAS ECAL) we find that
\begin{equation}
  \label{eq:2}
  E \ge \left\{
    \begin{array}{lcl}
     87\GeV&\text{for}& \pi^0\text{-meson}\\
      354\GeV&\text{for}&\eta\text{-meson}
    \end{array}
    \right.\;.
\end{equation}
This naive estimate suggests that for a sufficient number of isolated
$\eta$-meson a feature at $E \gtrsim 350\gev$ should appear in single photon
spectrum. Correspondingly, the diphoton invariant mass would have a similar
feature at twice this energy. Of course realistic photon reconstruction at
ATLAS/CMS is much more sophisticated~\cite{Aad:2010sp,Khachatryan:2015iwa} and depends on the details of the
detector. Below we take into account some of the factors: realistic energy and
angular meson distribution; geometry of the ECAL pixels; all neutral decay
modes of $\eta$ meson.

\begin{figure}
  \centering
  \includegraphics[width=0.5\textwidth]{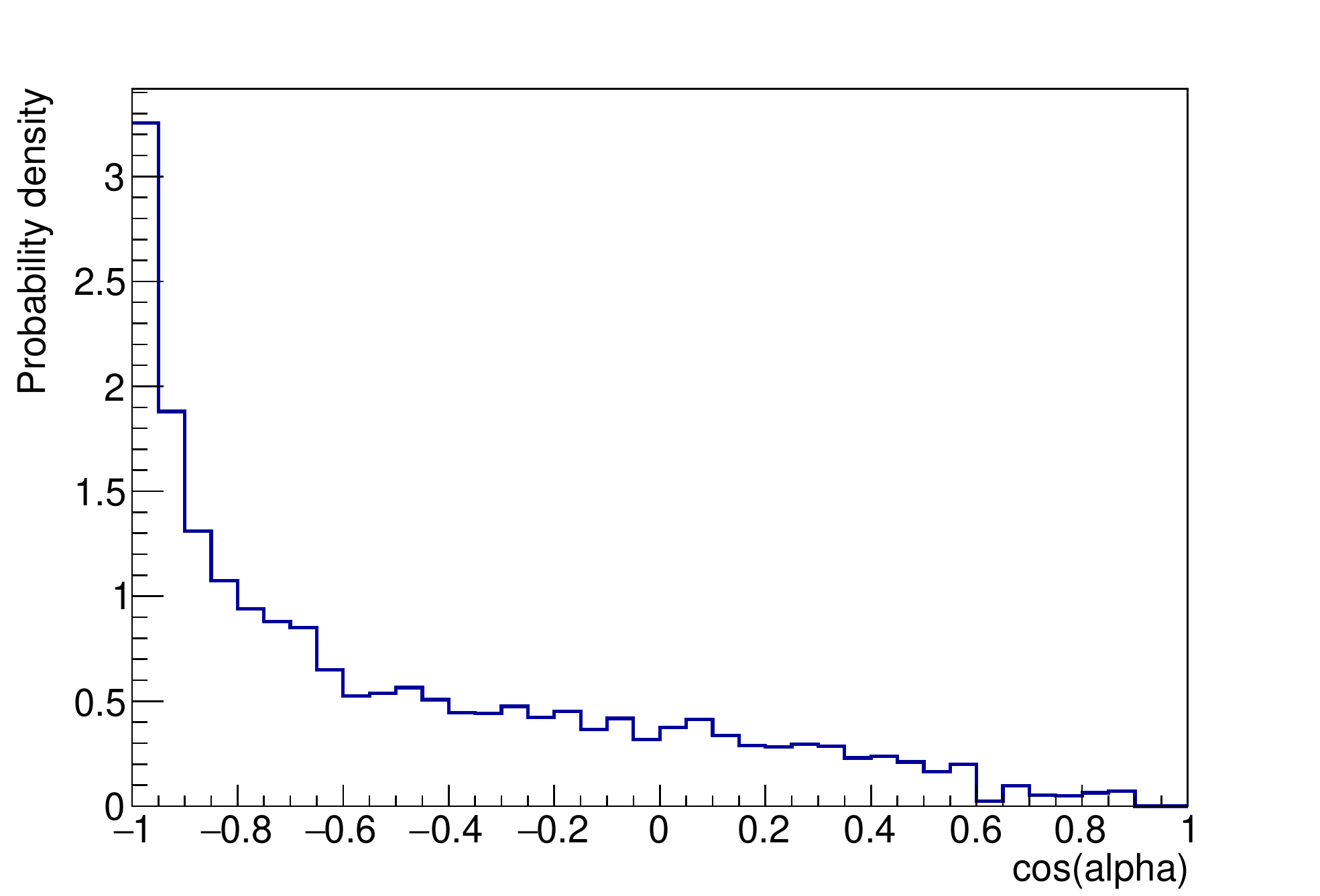}
  \caption{Distribution of angles between two $\eta$-mesons. A peak at
    $\cos\alpha = 1$ (corresponding to two mesons in the same jet) is not
    shown.}
  \label{fig:CosAngle}
\end{figure}

\subsection{Isolated neutral mesons} 

Using \textsc{pythia8} simulations~\cite{Sjostrand:2007gs} we find that the
number of $\eta$-mesons at energies of interest is well approximated by the
exponential
\begin{equation}
  \label{eq:3}
  \frac{dN_\eta}{dE} \propto e^{-E/{84\gev}}\;.
\end{equation}
We also find that in events where two $\eta$-mesons do not belong to a single
jet, they are mostly back-to-back. The distribution of angles between two
$\eta$-mesons (excluding $\cos \alpha = 1$ bin) is shown in
Fig.~\ref{fig:CosAngle}.

\subsection{Realistic energy and angular distribution of photons}  

The above
estimate~(\ref{eq:2}) is modified by the fact that the realistic pixel has a
different shape.  Namely, for photons that arrive aligned along the
$\phi$-direction (where resolution is much lower) such probability is non-zero
for low energies.  
In case of $\eta\to6\gamma$
  decay, the effect of all 6 photons aligned along the $\phi$ direction is
  drastically reduced and therefore one can think about the ATLAS pixel as
  having ``square'' form with both dimensions being $\Delta \eta_A$.

\section{Results: feature in di-meson invariant mass spectrum} 
Finally, we
generate the invariant mass, $m_{\eta\eta}$ of two $\eta$-mesons
misinterpreted as two photons.  To this end we perform the following procedure:
\begin{compactenum}[(1)]
\item Use the function~(\ref{eq:3}) as probability density function and draw
  from it two random energies of $\eta$-mesons, $E_1$ and $E_2$;
\item Use distribution of cosines between two $\eta$ mesons
  (Fig.~\ref{fig:CosAngle}) to generate random $\cos\alpha$;
\item We calculate invariant mass $m_{2\eta}$ as
  \begin{equation}
  m_{2\eta} = \sqrt{E_1 E_2 - p_1 p_2 \cos\alpha},\qquad p_i = \sqrt{E_i^2 - m_{\eta}^2}\;.
  \label{eq:5}
\end{equation}
\end{compactenum}
The result for the ATLAS is shown in Fig.~\ref{fig:dieta_mass_Energy100_ATLASsq_3pions_05}
where one can see that di-eta invariant mass spectrum has a peak at energy
$\sim 750\gev$. To make a definitive conclusion regarding the contribution of
this effect to the di-photon
feature~\cite{ATLAS-diphoton,CMS:2015dxe,ATLAS-diphoton2016}, we need to
determine the correct normalisation of the
peak in Fig.~\ref{fig:dieta_mass_Energy100_ATLASsq_3pions_05}, which one cannot
do without realistic Monte Carlo simulations.
The corresponding peak for CMS is
located at about $5.8$ times lower energies (the ratio of
$\Delta\eta_C/\Delta \eta_A$, see Table~\ref{tab:granulariites}), curiously
falling into the range of the Standard Model Higgs boson.
%

\begin{figure*}
  \centering
  \includegraphics[width=0.5\textwidth]{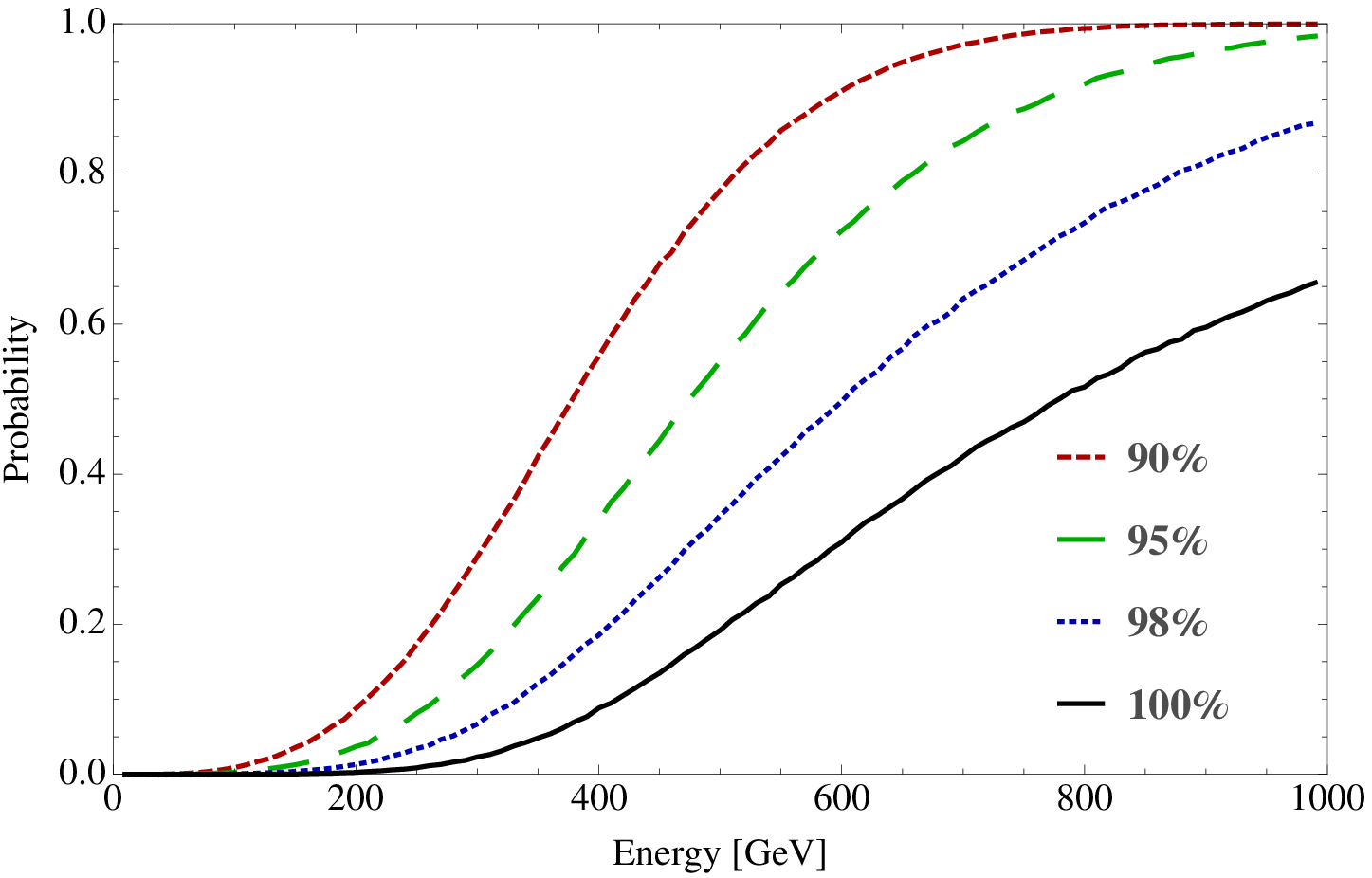}~
	\includegraphics[width=0.5\textwidth]{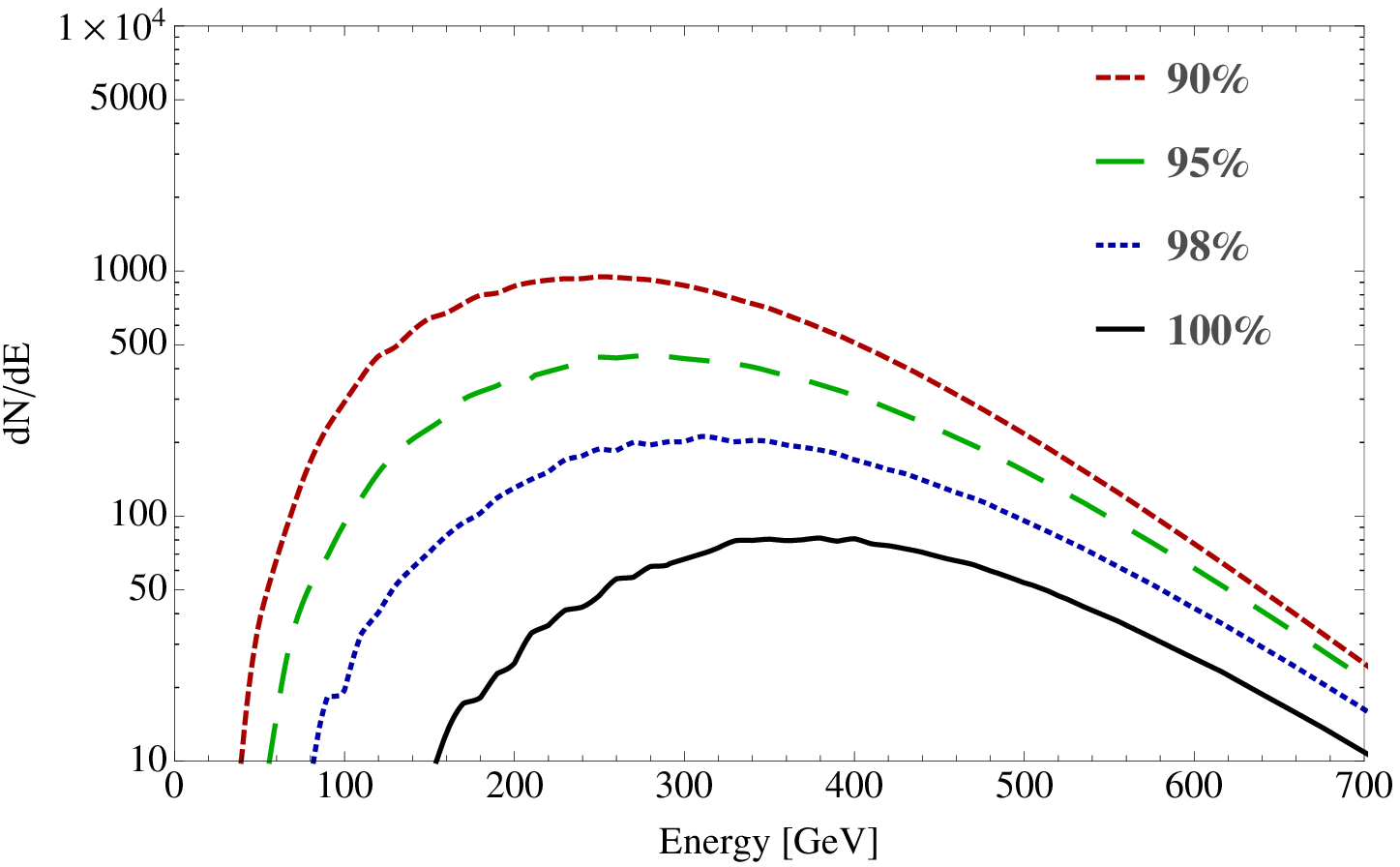} 
        \caption{\textit{Left:} Probability that a fraction of the total energy
          carried by some of $6\gamma$ is deposited in a single pixel. \textit{Right:}
          The same probability convoluted with Eq.~(\protect\ref{eq:3}).}
        \label{fig:EnergyX_ATLAS}
\end{figure*}

\section*{Discussion}

A localized exsess in the diphoton invariant mass spectrum is one of the
preferred ways to look for new particles due to its sufficiently low
background. The searches for new particles with diphotons implicitly assume
that the Standard Model background, as measured by the ATLAS or CMS detectors
is \emph{smooth} along the energies of interest.

In this work we investigated this assumption and demonstrated that there are
several potential factors that can produce bump-like features in the otherwise
smooth spectrum of diphotons in the Standard Model.  They are associated with
calorimeter granularity, experimental cuts, and single photon
misidentification.  Of course, we cannot really claim that these bumps are indeed seen in experiment or even that they could be seen at LHC at all, as this would require to make a number a number of checks of our hypothesis, listed below. 

\begin{enumerate}
\item As the neutral mesons are part of jets (probably carrying a large
  fraction of jet's $p_T$), making stronger photon isolation cuts should
  decrease the excess (while it should not affect the actual
  physical diphoton signal of course).

\item The region around $E\sim 350-400\gev$ in single photon (photon + jet)
  spectrum may reveal a feature if our hypothesis is correct.

\item Finally, the best way to check this hypothesis is to perform the
  diphoton analysis  over a Standard Model Monte Carlo
  simulations, using full detector simulation and applying the same
  types of cuts as in~\cite{ATLAS-diphoton,CMS:2015dxe},
\end{enumerate}

Clearly, our results are rather qualitative, as the detector responses, in
particular the isolation requirements used in the experimental analysis for
suppression of jet background, were very crudely modeled in this work. First
of all, to estimate the size of this effect (the total number of events in the
``excess'') we need to know the absolute number of isolated $\eta$-mesons at
energies of interest. We understand that this number is tiny which makes it
more difficult to estimate. Additional details of photon reconstruction also
affect shape and position of the bump (and our Fig.~\ref{fig:EnergyX_ATLAS}
illustrates this).

Even if the 750~GeV excess will not be confirmed with more data or with
refined analysis, it is important to understand whether the observed excess
was the fluctuation or unaccounted background contribution as our note
suggests. As the searches in the diphoton channel will continue, clarifying
the exact shape of the Standard Model diphoton background will remain an
important question.

A final remark.  None of the authors of this paper is an expert in detector
physics, nor do we  have an access to details of the experimental analysis or
to large scale Monte-Carlo simulations which can estimate the $\eta$ production. Still, we find it quite curious that the appearing bump energy scale is in the interesting region around 750~GeV, and that the $\eta$ decays (to the best of our knowledge) were not discussed in this connection. These considerations lead us to idea to make this note public.

\myparagraph{Acknowledgements.}  The work of MS was partially supported by the
Swiss National Science Foundation. We would like to thank L.~Shchutska, P.~Hansen and
S.~Xella for help with understanding the details of ATLAS and CMS ECALs and MC simulations.

\bibliographystyle{JHEP-2} %
\bibliography{750gev} %

\end{document}